\documentclass[prd]{revtex4}
\usepackage{epsfig}

\usepackage{graphicx}

\begin{document}


%
%

\title{BEPCII and BESIII}

\author{Frederick A. Harris \\
(For the BES Collaboration)}
\address
{Dept. of Physics and Astronomy, \\
        The University of Hawaii, \\
        Honolulu, Hawaii 96822, USA \\
fah@phys.hawaii.edu}




\begin{abstract}
The Beijing Electron Collider has been upgraded (BEPCII) to a two-ring
collider with a design luminosity of $1 \times 10^{33}$cm$^{-2}$
s$^{-1}$ at a center-of-mass energy of 3.78 GeV. It will operate
between 2 and 4.6 GeV in the center of mass.  With this luminosity,
the BESIII detector will be able to collect, for example, 10 billion
$J/\psi$ events in one year of running.  This will be a unique
facility in the world opening many physics opportunities. BEPCII and
BESIII are both currently being commissioned, first events have
been obtained, and data taking will take place in fall 2008.

\end{abstract}


\maketitle

\section{Introduction}

The Beijing Electron-Positron Collider (BEPC) at the Institute of High
Energy Physics (IHEP) in Beijing was, until CLEOc in 2003, a unique
facility running in the tau-charm center-of-mass energy region from 2
to 5 GeV with a luminosity at the $J/\psi$ peak of $5 \times
10^{30}$cm$^{-2}$ s$^{-1}$.  The Beijing Spectrometer
(BESI~\cite{besi} and BESII~\cite{besii}) detectors at the BEPC had
operated since about 1990 and studied many physics topics, including a
precision measurement of the tau mass~\cite{taumass} and a detailed
R-scan~\cite{rscan}, and obtained 58 million events at the $J/\psi$,
14 million at the $\psi^{'}$, and over 30 pb$^{-1}$ at the $\psi{''}$.

 In 2003, the Chinese Government approved the upgrade of the BEPC to a
two-ring collider (BEPCII) with a design luminosity approximately 100
times higher than that of the BEPC.  This will allow unprecedented
physics opportunities in this energy region and contribute to
precision flavor physics.  In this paper, BEPCII and BESIII will be
described, along with their status. Additional details may be found in
previous ones~\cite{bes3a}.


\section{BEPCII}

BEPCII is a two-ring $e^+e^-$ collider that will run in the tau-charm
energy region ($E_{cm} = 2.0 - 4.2$ GeV, but possibly as high as 4.6
GeV) with a design luminosity of $1\times 10^{33}$ cm$^{-2}$s$^{-1}$
at a beam energy of 1.89 GeV, an improvement of a factor of 100 with
respect to the BEPC. This is accomplished mainly by using
multi-bunches and micro-beta.

\begin{figure}  \centering
   \includegraphics*[width=0.55\textwidth]{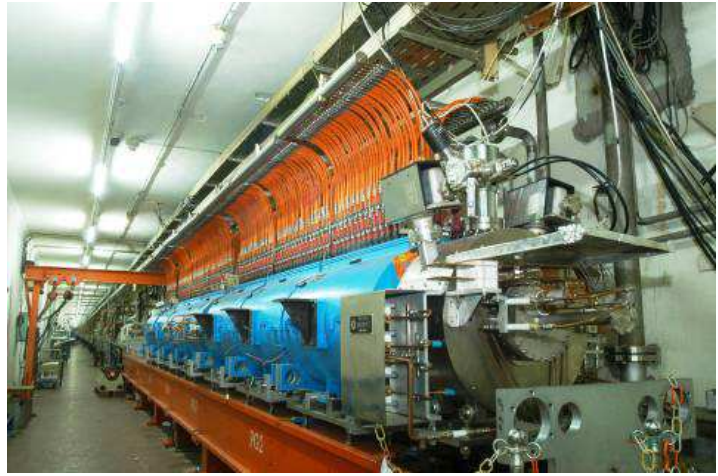}  
  \caption{\label{linac}The completed LINAC of BEPCII.
    }
 \end{figure}

The 202.4 meter long linac has been upgraded with new klystrons, a new
electron gun, and a new positron source to increase its energy and
beam current; it can accelerate electrons and positrons up to 1.89 GeV
with an positron injection rate of 62 mA/min. Its installation was
completed in the summer of 2005 (see Fig.~\ref{linac}), and it has
reached or surpassed all design specifications.

There are two storage rings with lengths of 237.5 meters.  The
collider has new super-conducting RF cavities, power supplies, and
control; super-conducting quadrupole magnets; beam pipes; magnets and
power supplies; kickers; beam instrumentation; vacuum systems; and
control.  The old dipoles are modified and used in the outer ring.
Electrons and positrons collide at the interaction point (IP) with a
horizontal crossing angle of 11 mrad and bunch spacing of 8 ns.  Each
ring has 93 bunches with a design beam current of 910 mA.  The machine
also provides a high flux of synchrotron radiation at a beam energy of
2.5 GeV.  

Commissioning of the new collider with detector installed is in progress.
So far, a luminosity of $1 \times 10^{32}$ cm$^{-2}$s$^{-1}$ and beam currents of 550 mA
for both beams have been achieved.

\section{BESIII}
\begin{figure}  \centering
   \includegraphics*[width=0.55\textwidth]{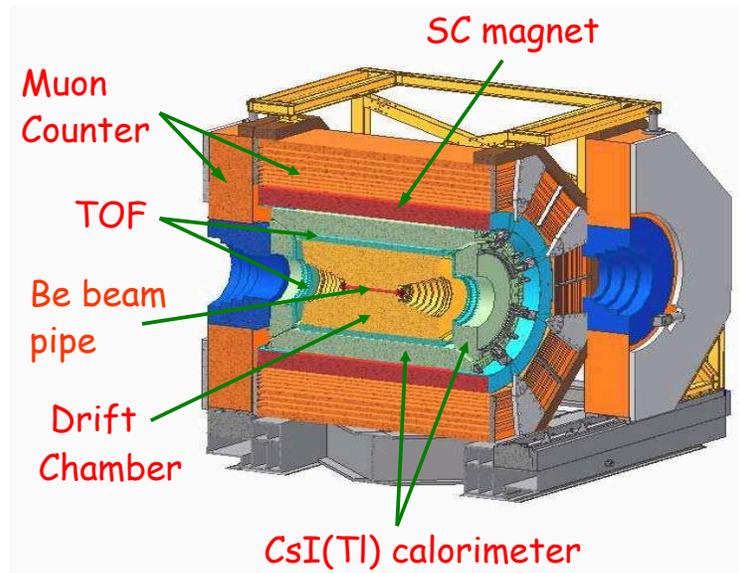}  
  \caption{\label{schematic}Schematic view of the BESIII detector.
    }
 \end{figure}

The BESIII detector consists of a beryllium beam pipe, a helium-based
small-celled drift chamber, Time-Of-Flight counters (TOF) for particle
identification, a CsI(Tl) crystal calorimeter, a super-conducting
solenoidal magnet with a field of 1 Tesla, and a muon identifier using
the magnet yoke interleaved with Resistive Plate Counters
(RPC). Fig.~\ref{schematic} shows the schematic view of the BESIII
detector, including both the barrel and end cap portions.
    
\subsection{Main Drift Chamber}    
\begin{figure}  \centering
   \includegraphics*[width=0.55\textwidth]{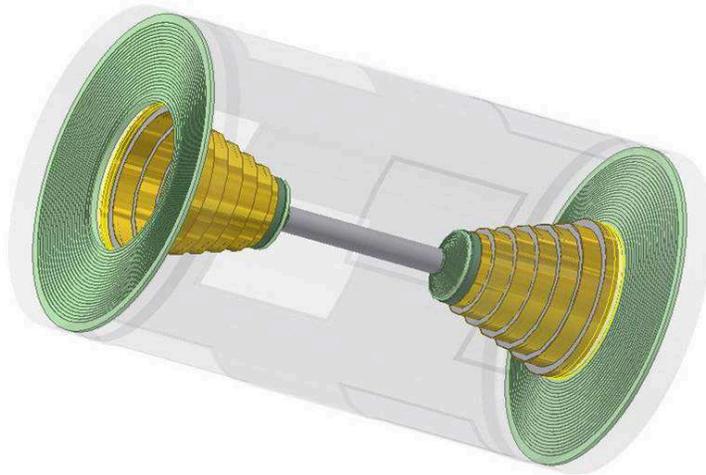}  
  \caption{\label{mdc}Schematic view of the MDC.
    }
 \end{figure}

The main drift chamber (MDC) is 2.58 meters in length and has an inner
radius of 59 mm and an outer radius of 0.81 m.  The inner and outer
cylinders are carbon fiber. As shown in Fig.~\ref{mdc}, there is a
short inner portion near the beam pipe, a stepped region, and a cone
shaped outer region.  The polar angle coverage is $\cos \theta = 0.83
$ for a track passing through all layers, and $\cos \theta = 0.93 $
for one that passes through 20 layers.  The end-plates are machined
with a hole position accuracy better than 25 microns. Altogether there
are 43 layers of 25 micron gold plated tungsten sense wires; the field wires
are 110 micron gold-plated aluminum.  The cells are approximately
square, and the size of the half-cell is 6 mm in the inner portion of
the drift chamber and is 8.1 mm in the outer portion.  The chamber
uses a 60/40 He/$C_3H_8$ gas mixture.

The design spatial, momentum, and $dE/dx$ resolutions are $\sigma_s =
130 \mu$m, $\sigma_p/p= 0.5 \% $ at 1 GeV/$c$, and
$\sigma_{dE/dx}/dE/dx \sim 6 \%$, respectively.  Beam tests performed
with prototype electronics at KEK in a 1 T magnetic field yielded a
spatial resolution better than 110 microns and $dE/dx$ resolution
better than 5\%. Tests of the final chamber and readout electronics
using cosmic rays without magnetic field yield a spatial resolution
of $\sigma_s = 139 \mu$m and cell efficiency of 97 \%; better
resolution is expected with colliding beam data and further
calibration.  The readout uses the CERN HPTDC.

\subsection{TOF}

Outside the MDC between a radius of 810 mm and 930 mm is the time of
flight (TOF) system, which is crucial for particle identification. It
consists of a two layer barrel array of 88 50 mm x 60 mm x 2380 mm
BC408 scintillators in each layer and end cap arrays of 48 fan shaped
BC404 scintillators.  Hamamatsu R5942 fine mesh photo-tubes are used -
two on each barrel scintillator and one on each end cap scintillator.
Expected time resolution for kaons and pions and for two barrel layers is 100
to 110 ps, giving a $2 \sigma$ $K/\pi$ separation up to 0.9 GeV/c for
normal tracks.  This resolution has been confirmed in a beam test of a TOF
counter using prototype electronics.  A TOF monitoring system
featuring a 440 nm laser diode and specially designed fiber optic
bundles was built by the University of Hawaii, IHEP, and the
University of Science and Technology of China~\cite{tofmon}.

\subsection{Calorimeter}
\begin{figure}  \centering
   \includegraphics*[width=0.30\textwidth]{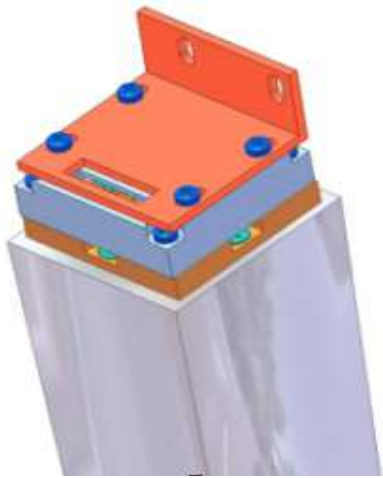}  
  \caption{\label{crystal}Schematic of the photo-diode and preamp
    assembly on the end of an electromagnetic calorimeter crystal.
    }
 \end{figure}

The CsI(Tl) crystal calorimeter contains 6240 crystals total in the
barrel and end cap portions of the calorimeter.  The typical
crystal is 5.2 $\times$ 5.2 cm$^2$ on the front face and 6.5 $\times$ 6.5
cm$^2$ on the rear face with a length of 28 cm or 15 radiation
lengths.  Figure~\ref{crystal} shows a schematic of the assembly
containing an aluminum plate with two photo-diodes (Hamamatsu S2744-08)
with 10 mm by 20 mm sensitive area and an aluminum box for the two
preamps and amplifier
mounted on the end of a crystal.  The design energy and spatial
resolutions at 1 GeV are 2.5 \% and 0.6 cm, respectively, and the
energy range will extend as low as 20 MeV.

\begin{figure}  \centering
   \includegraphics*[width=0.55\textwidth]{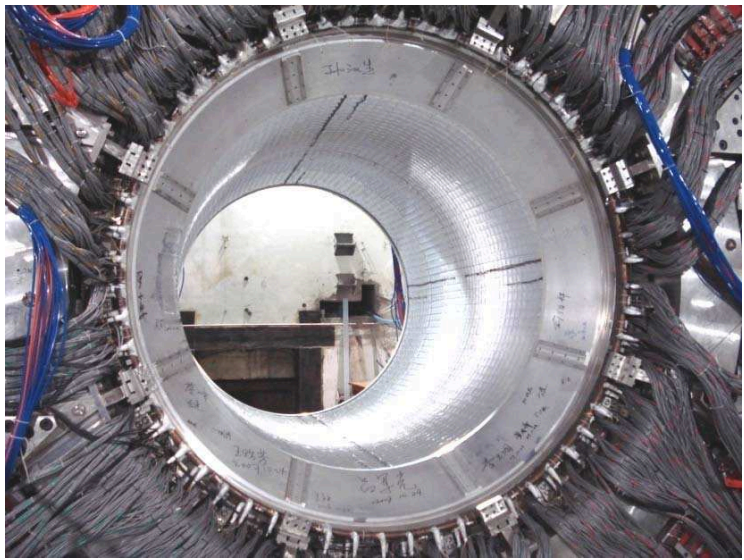}  
  \caption{\label{calorimeter} Assembled barrel electromagnetic
    calorimeter inside the superconducting magnet.
    }
 \end{figure}

Figure~\ref{calorimeter} shows barrel electromagnetic calorimeter
inside the superconducting magnet.  The crystals are held by screws
and there are no walls between crystals.

\subsection{Magnet}
\begin{figure}  \centering
   \includegraphics*[width=0.55\textwidth]{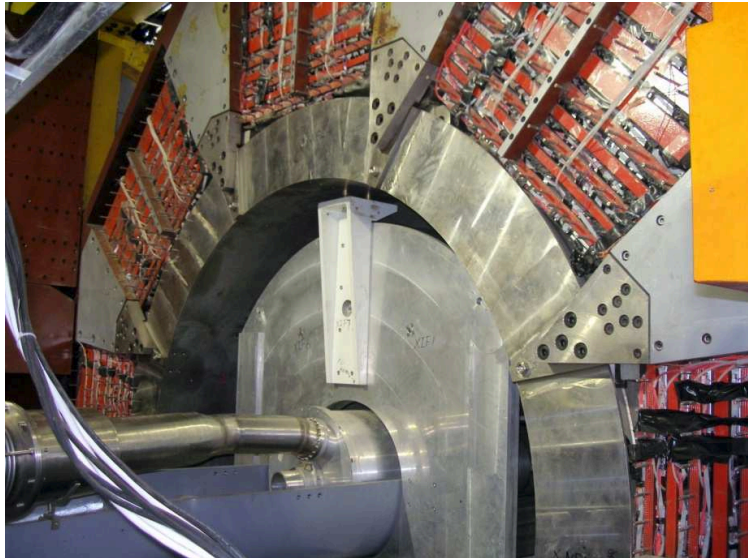}  
  \caption{\label{magnet}
The magnet during 
field mapping, done with the super conducting quadrupoles in
place, using a computer controlled mapping machine during
2007. Outside the magnet, the return steel and RPC chambers of the
muon counter are visible.}
 \end{figure}

The BESIII super-conducting magnet is the first of its kind built in
China.
It is a 3.91 m long
single layer solenoid with a 1 T magnetic field at a nominal current
of 3369 A. 
Fig.~\ref{magnet} shows the magnet during 
field mapping in June 2007, done with the super conducting quadrupoles in
place, using a computer controlled mapping machine.
The measured field in the MDC has an accuracy better than 0.25 \% and was
measured with a position accuracy of 0.5 mm.

\subsection{Muon Counter}

The magnet return iron has nine layers of Resistive Plate Chambers (RPC)
in the barrel and eight layers in the end cap to form a muon counter,
which can be seen in Fig.~\ref{magnet}.  The
electrodes of the RPCs are made from a special phenolic paper laminate
on Bakelite, which has a very good surface quality. The gas used is Ar
: $C_2H_2F_4$ : Isobutane (50:42:8).  Extensive testing
and long term reliability testing have shown that the chambers have
high efficiency, low dark current, and good long term stability.
The RPCs use 4 cm wide one dimensional readout strips, and about 10,000
channels of readout are required.

    

\subsection{Trigger, Data Acquisition, and Offline Software}

The trigger is pipelined and uses
FPGAs.  Information from the TOF, MDC,  and muon counter are
used. The maximum trigger rate at the $J/\psi$ will be about 4000 Hz
with a good event rate of about 2000 Hz. 

The whole data acquisition system has been tested to 8 kHz for an
event size of 12 Kb, which is a safety margin of a factor of two.  The
expected bandwidth after level one is 50 Mbytes/s. The data
acquisition system has 1000 times the performance of BESII.
    
The BES Offline Software System (BOSS) is complete.
Simulation is based on Geant4. 
    


\subsection{Status}

In March 2008, a two month long cosmic ray run without magnetic field
was completed.  This run was extremely useful for commissioning the
detector and doing initial calibrations and performance checks.  

\begin{figure}  \centering
   \includegraphics*[width=0.55\textwidth]{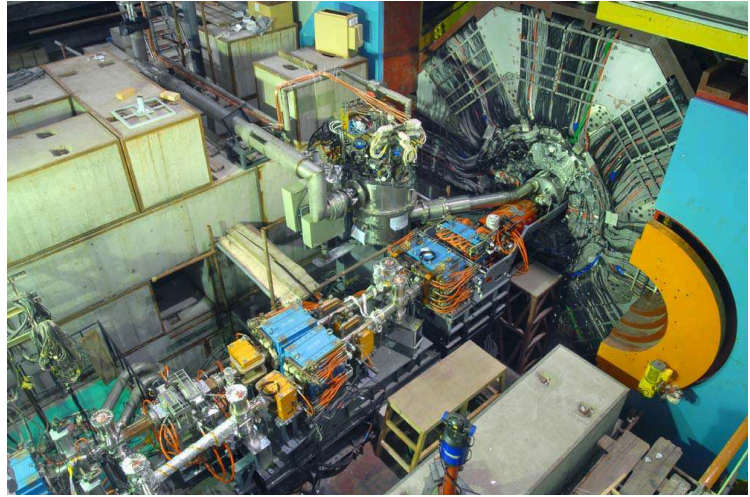}  
  \caption{\label{besatIP}BESIII detector at the IP in June
  2008. Shown are a superconducting quadrupole, the two beam lines,
  and the BESIII detector with the magnet iron open and end caps
  exposed.  }
 \end{figure}

The detector then moved to the IP and is shown in Fig.~\ref{besatIP}
at its final location in June 2008 with all beam magnets and vacuum
pipes in place.

\begin{figure}  \centering
   \includegraphics*[width=0.55\textwidth]{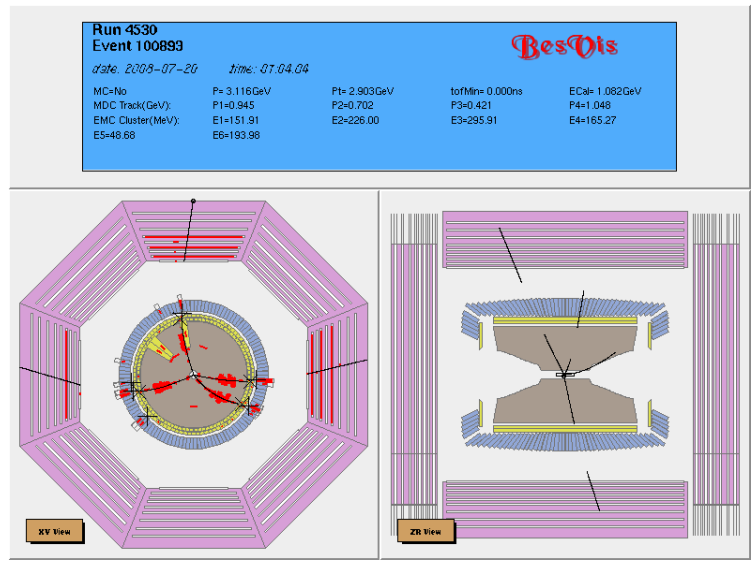}  
  \caption{\label{1stevent}Online display of the first hadronic event
  recorded by the BESIII detector on July 19, 2008.}
 \end{figure}

Commissioning of the detector and collider together began in July,
and the first hadronic event was obtained on July 19, 2008 (see
Fig.~\ref{1stevent}).  Data taking will begin in early fall. 

The BESIII collaboration includes physicists from IHEP, many Chinese
Universities, and groups from Germany, Japan, Russia, and the U.S.

\section{Physics in the tau-charm energy region}

The tau-charm energy region makes available a wide variety of
interesting physics.  Data can be taken at the $J/\psi$, $\psi(2S)$,
and $\psi(3770)$, at $\tau^+\tau^-$ threshold, and at an energy to allow
production of $D_S$ pairs, as well as for an R-scan.  The $\psi(3770)$
is an ideal factory for producing $D \bar{D}$ pairs, and the
$\psi(2S)$ allows access to $\eta_c$, $h_c$, and $\chi_c$ physics via
radiative and hadronic transitions.

\begin{table}[htb]
\caption{
 Number of events expected for one year of running.}
{\begin{tabular}{| l | c | c | c | c| }
\hline
Physics & Center-of-mass  &  Peak          & Physics      & Number of \\ 
channel & Energy          & Luminosity     & cross    & Events per \\ 
        & (GeV)  & ($10^{33}$ cm$^{-2}$ s$^{-1}$) & section (nb) & year\\\hline
$J/\psi$   &   3.097  &  0.6  &  $\sim 3400$ & $10\times 10^9$ \\
$\tau$     &   3.67   &  1.0  &  $\sim2.4$   & $12 \times 10^6$ \\
$\psi(2S)$ &   3.686  &  1.0  &  $\sim640$   & $3.0 \times 10^9$ \\
$D$        &   3.770  &  1.0  &  $\sim5$     & $25 \times 10^6 $\\
$D_S$      &   4.030  &  0.6  &  $\sim0.32$  & $1.0 \times 10^6$ \\
$D_S$      &   4.140  &  0.6  &  $\sim0.67$  & $2.0 \times 10^6$ \\
\hline \end{tabular}\label{3.11}}
\end{table}

Clearly BESIII with higher a luminosity of $1 \times 10^{33}$
cm$^{-2}$ s$^{-1}$ will contribute greatly to precision flavor physics;
$V_{cd}$ and $V_{cs}$ will be measured with a statistical accuracy of
better than 1.0\%. $D^0 \bar{D^0}$ mixing will be studied, and CP violation
will be searched for. Table~\ref{3.11} gives the numbers of events
expected during one year of running at various energies.  Huge
$J/\psi$ and $\psi(2S)$ samples will be obtained.  The $\eta_c$,
$\chi_{cJ}$, and $h_c$ can be studied with high statistics.  The $\rho
\pi$ puzzle will be studied with better accuracy.  The high statistics
will allow searches for physics beyond the standard model.  The future
is very bright.

\section{Acknowledgments}
It is important to acknowledge all the hard work by personnel at IHEP
and by BESIII collaborators on both BEPCII and BESIII.  This work was
supported by the Department of Energy under Contract No.
DE-FG03-94ER40833 (U Hawaii).




\end{document}